\documentstyle[11pt,newpasp,twoside]{article}
\def\edcomment#1{\iffalse\marginpar{\raggedright\sl#1\/}\else\relax\fi}

\reversemarginpar

\begin{document}

\title{Concluding Remarks}
\author{P. J. E. Peebles}
\affil{Joseph Henry Laboratories, Princeton University,
Princeton, NJ 08544, USA}

 
\begin{abstract}
I review the reason for considering the prime purpose of the
program of measurements of the fundamental parameters of
cosmology to be the tests of cosmological models. I comment on the 
philosophy by which we are approaching this goal, offer an
assessment of where we stand, and present some thoughts on where
the tests may be headed.  
\end{abstract}

\section{Introduction}

These Proceedings document impressive progress toward a
satisfactory completion of the great program of measurements of
the parameters of cosmology that commenced in the 1930s. It has
taken a long time, and has brought into play many phenomena
and measurements that could not have been anticipated in the
1930s. We may at last be approaching closure of this program, and
it is appropriate to reflect on why we are so interested in these 
measurements and how it informs our interpretation of the
results.   

\section{The Significance of the Cosmological Tests}

I take the literal reading, that the purpose of the 
cosmological tests is to test models, in particular
the commonly accepted relativistic Friedmann-Lema\^\i tre
cosmology. It certainly is useful to have byproducts such as the
demonstration of the presence of a term in the  
stress-energy tensor that acts like Einstein's cosmological
constant $\Lambda$, which may help guide us to a resolution of
the perplexing physics of the energy density of the vacuum, and
a measurement of the radius of curvature of space
sections at constant world time, which may prove to be a clue
to what the universe was like before it could have been described
by the Friedmann-Lema\^\i tre model. But all this is true only
if we have convincing reason to trust the basis for these
results. 

A part of cosmology we can trust is the near homogeneous
evolution of the observable universe from a denser hotter state. 
The list of evidence is familiar but worth repeating to 
make the point: we have compelling reason to believe this
is what happened. Deep counts of objects at wavelengths ranging
from radio to gamma rays are close to isotropically distributed
across the sky. Either we are close to a center of spherical
symmetry or our universe is close to homogeneous. If the latter,
and the distribution is expanding so as to preserve homogeneity
and isotropy, the recession velocity satisfies Hubble's law. The
low redshift part of the SNeIa measurements is an
impressively tight demonstration of the redshift-distance
relation. The cosmological interpretation of quasar redshifts
passes demanding tests, such as the tight 
correlation of Lyman-limit and Mg~II absorption lines with
galaxies at close to the same angular position and redshift,
showing quasars are behind lower redshift
galaxies. If the expansion traces back to very high density
galaxies at high redshift are seen as they were closer to the 
time when galaxies could not have existed, and ought to look
younger than nearby ones. The effect is amply demonstrated. The
3~K radiation (the CBR) could not have relaxed to its
thermal spectrum in the universe as it is now because space is
not opaque at the Hubble length: radio sources are observed at
$z\sim 1$. We can understand the thermal 
spectrum if the universe has expanded from a denser, hotter state
that is optically thick within the Hubble length. The angular
position of the peak of the spectrum of angular fluctuations of
the CBR agrees with the conventional physics of the evolution of
primeval adiabatic mass density fluctuations if the universe has
expanded and cooled by a factor much larger than 
$z_{\rm eq}\sim 1000$, the redshift of decoupling of matter and
radiation. Helium and deuterium are natural byproducts of
expansion from still higher temperature. 

This list offers no guidance to what happened at
very large redshift, or well outside the Hubble length. The
inflation concept has shown us how easy it is to imagine the
universe at great distance is not at all like what we see, 
but determining whether such an ``island universe'' picture is  
realistic is outside the current round of cosmological tests as I
would define them. I don't know whether the people in the 1930s
who pioneered the program of cosmological tests gave much thought
to the spatial and temporal limitations of empirical evidence
within their cosmology. If not we have to adjust the program.  

You can add to the list of evidence for evolution, depending on 
how much you want to rely on models, but I think the point is
clear: it employs a broad variety of phenomena
observed in quite different ways. Individual entries could be
wrong, but it would be absurd to imagine all quite consistently 
point in the wrong direction. Thus Hoyle, Burbidge \&\ Narlikar
(1993) accept cosmic evolution, but argue the last substantial
addition to the entropy in the CBR could have occurred at a much
more modest expansion factor than in the standard model. This 
is a considerable difference, but it should not obscure the
point that the redundancy of evidence has forced us to the answer
to Hubble's (1936) question, is the cosmological redshift the
result of the general recession of the nebulae? It is, and the
general recession is associated with cosmic evolution.  

Our answer to Hubble depends on local physics and symmetry
arguments, but it makes little use of general relativity theory
(hereinafter GR); I did not even mention the relativistic
relations among observables that Tolman (1934)
listed.\footnote{The theory of the origin of the light elements
assumes the expansion rate equation, $H^2=8\pi G\rho /3$. This
follows from local physics; relativity enters only in the
expression for active gravitational mass. The CBR anisotropy
computation uses the 
angular size distance-redshift relation, a highly nontrivial
application of the large-scale spacetime geometry, but the
success of the prediction forces its inclusion in my list of
elementary evidence for cosmic evolution.} The observations of
supernovae of type Ia  probe one of the relations, between
magnitude and 
redshift, and detect a departure from the Einstein-de~Sitter
case. This magnificent accomplishment is in no way depreciated by 
noting that by itself it is not a cosmological test. In addition
to the slight chance some quirk of the physics of supernovae has 
avoided the thorough checks for systematic error (or, to
be really cautious, that Nature has put us at the
center of a spherically symmetric universe with a slight radial
density gradient), the conventional interpretation depends on GR,
and, within this theory, the measurement is readily fitted by the
adjustment of free parameters. Hoyle, Burbidge \&\ 
Narlikar (1993) might similarly fit the measurement by suitable
choice of parameters within their theory.   

The elegant logic of general relativity theory, and its precision
tests, recommend GR as the first
choice for a working model for cosmology. But the Hubble
length is fifteen orders of magnitude larger than the length
scale of the precision tests, at the astronomical unit and
smaller, a spectacular extrapolation. The extrapolation is tested
by checking for consistency of the cosmological parameters
derived from different aspects of the geometry of spacetime. The
Robertson-Walker line element figures in Tolman's (1934) list of
cosmological relations. The computation of the CBR anisotropy
spectrum uses GR to propagate the irregularities in the radiation 
distribution through spacetime that is predicted to be strongly
curved over the expansion factor $z\sim 1000$ since decoupling,
and it uses GR to predict the dynamics of small fluctuations in
the distributions of matter and radiation at $z\sim 1000$. The
dynamical estimates of galaxy masses from rotation curves and
streaming velocities assume the latter aspect of GR, the
inverse square force law for gravity. Weak and strong
gravitational lensing use this law, with the usual factor of two
correction. A tight check of consistency of the parameters
derived from these different phenomena would be a demanding test
of GR and the cosmology. 

The spectrum of angular fluctuations of the CBR offers a
wonderfully rich basis for these tests. These Proceedings discuss
the constraints on the density parameters in
dark matter and in Einstein's cosmological constant $\Lambda$ (or
a term in the stress-energy tensor that acts like $\Lambda$), to
be compared to what is indicated by the dynamical measurements of
masses of galaxies and systems of galaxies, by the curvature of
the redshift-magnitude relation, and by the measurement of
$H_ot_o$; the density 
parameter in baryons, to be compared to the theory and
observational tests of the origin of helium and deuterium at high 
redshift and to the observational baryon budget at low redshift;
the density parameter in neutrinos, to be compared to laboratory
and atmospheric oscillation experiments; and the amplitude
of the primeval density fluctuations, to be compared to
measurements of the distributions of galaxies and mass
at low redshift. 

The impressive consistency of constraints that have already
emerged from
such different applications of GR and the cosmological principle
suggests the theory and the cosmology are on the right track. We
should be cautious about the details, however, because the
interpretation of the CBR anisotropy also assumes the adiabatic
cold dark matter (CDM) theory for structure formation, and the
tests of this model depend on some subtle issues of astronomy.  

\section{The Model for Structure Formation and the Issue of Voids}

We pay particular attention to simple and elegant ideas in
physical science because Nature tends to agree with us. 
We have examples in cosmology: GR, Einstein's cosmological
principle, and the adiabatic CDM model for structure
formation.\footnote{One can think of many other 
arguably less elegant models for structure formation; you can
trace through astro-ph my list of alternatives, each killed
by the inexorable advance of the measurements.} But Nature 
is quite capable of surprising us, as witness the evidence for a
significant cosmological constant, which a few years ago
was generally considered to have no socially redeeming value.
Since many of the cosmological tests depend on the CDM model we
must consider its empirical tests.

In these Proceedings Carlos Frenk and John Peacock present 
impressive observational successes of the CDM model. There are a
few clouds on the small-scale part of the horizon, however; an
example that has particularly impressed me is the void phenomenon
(Peebles 1989).
Carignan and Freeman's (1988) ``dark galaxy,'' DDO~154, seems to
be a close approximation to one of the failed galaxies that
figure in commonly discussed interpretations of numerical
simulations of the CDM model. In these simulations there is
appreciable mass in the voids defined by the positions of dark mass
concentrations that are massive enough to qualify as homes for
normal $L\sim L_\ast$ high surface brightness galaxies. This void
medium contains low mass halos that would seem to be acceptable
homes for galaxies like DDO~154. So why are galaxies like DDO~154
not found in the voids? 

There are void galaxies; nearby examples are the pair
NGC~6946 and NGC~6503. The former is an Arp (1966) peculiar
galaxy, but only because a supernova was seen in it. Sandage and
Bedke (1988) give 
a magnificent image of this galaxy; I have been assured it looks
like an ordinary large near face-on spiral, though maybe
unusually gas-rich. The other appears to be an edge-on
spiral; it is classified as Scd in the {\sl Nearby Galaxies
Catalog} (Tully 1988). The CDM model simulations
show occasional substantial upward mass fluctuations in generally
low density void regions that could be homes for $L\sim L_\ast$
void galaxies, but how would the baryons in these isolated
mass peaks get spun up to form normal-looking if isolated spirals?

To me the most remarkable and challenging phenomenon is that
observable objects respect the same voids. This
applies to giant and dwarf galaxies, and low and high surface
brightness ones (eg. Pustil'nik et al. 1995; Popescu, Hopp \&\
Rosa 1999; and references therein); to gas 
clouds observed in emission (eg. Zwann et al. 1997); and to high  
surface density gas clouds observed in absorption (Lanzetta et
al. 1995; Steidel, Dickinson \&\ Persson 1994).\footnote{Shull,
Stocke \&\ Penton (1996) show that gas clouds detected as very low
surface density Lyman~$\alpha$ absorbers avoid dense galaxy
concentrations. My impression is that they also avoid the voids,
but that is a subject of work in progress by Shull and colleagues.}

A common and defensible opinion is that the astrophysics by 
which void matter becomes visible as a galaxy of stars or an HI
or MgII absorber is so complicated as to quite confuse the 
interpretation of void phenomena. Cen \&\ Ostriker (2000) give 
an example: in their physically motivated prescription for
galaxy formation the void probability for all galaxies
identified in a simulation is much larger than for the mass.
Cen \&\ Ostriker conclude observed voids are not an argument
against CDM-like models. This is a valuable example of the
subtlety of the astrophysics, but I 
am even more impressed by the presence of dwarf galaxies on the
outskirts of the Local Group, isolated enough to seem to be
primeval rather than products of physical processes operating
within the large galaxies. These dwarfs are visible; why
should similar primeval halos in the voids be so cunningly
hidden?   

If void phenomena ruled out the CDM model we could turn to
alternatives. Bode, Ostriker \&\ Turok (2000) show that if the
CDM is replaced by warm dark matter it greatly reduces the
numbers of small dark mass halos, tends to produce dwarfs at
lower redshift, and yields smooth patches of dark matter within
the voids outlined by the massive halos. All are positive changes 
from CDM. But their figures~4 and~5 show caustics of dark matter
threading the voids. If these caustics fragmented into low mass
halos would the model predict greater numbers of dwarf or
irregular galaxies extending into the voids than is observed? If
the caustics remained smooth would the model predict more void
absorption line systems than is observed? It looks like a serious
challenge. 

I don't consider the void issue a very serious challenge to the 
Friedmann-Lema\^\i tre cosmology. Maybe I'm fooled by the
astrophysics, as Cen 
\&\ Ostriker (2000) argue. Maybe the CDM model must be adjusted,
perhaps along the lines of Bode, Ostriker \&\ Turok (2000),
perhaps in some other way. The magnificent prediction 
of the measured first peak of the CBR angular fluctuation spectrum 
shows the CDM model very likely is close to the
right picture. It would be less surprising to learn an improved 
structure formation model yields somewhat different constraints
on the cosmological parameters, of course. Here is a 
worked example. Suppose at $z\sim 1000$ there were objects with 
strong Lyman~$\alpha$ emission lines, like quasar spectra with
suppressed ionizing radiation; maybe primeval black holes. The
Lyman~$\alpha$ photons would delay recombination, preserving the
height of the first peak of the CBR fluctuation spectrum, but
shifting it to a larger angular scale for given cosmological
parameters, and biasing this measure of space curvature 
(Peebles, Seager \&\ Hu 2000).

\section{Is Cosmology a Science?}

Disney (2000) asks whether cosmology ``is a science at all,''
while I have been presenting it as a healthy and productive 
quantitative physical science. We might get some insight into the  
origin of these very different assessments from two
considerations.     

First, our commonly accepted cosmology did grow by the
introduction of hypotheses to fit phenomena. Some hypotheses have
been checked and established, as cosmic evolution. Some are being
checked, as dark matter. The dynamical mass estimates quite
consistently indicate the cosmological density parameter is low,
$\Omega _m\sim 0.2$. The SNe redshift-magnitude relation and the
measurement of $H_ot_o$ both favor this low value of $\Omega _m$.
As discussed in the last section we are assuming GR, but applying 
it in quite different ways. The consistency at the level we now
have is an elegant though not yet very precise test of GR and the
dark matter hypothesis. In short, cosmology does depend on
hypotheses, but we have nontrivial progress in testing them. 

Examples of work in progress are worth listing as a reminder of
how broadly based the cosmological tests are becoming. Consider
the projects to measure the predicted secondary peaks of the CBR 
temperature fluctuation spectrum and give us a first look at  
the polarization anisotropy, to test delayed recombination among
other things; measure the shape of the
redshift-magnitude relation at redshifts well above unity, to 
test the prediction that the expansion becomes matter-dominated;  
establish the constraint on parameters from the rate of
gravitational lensing of background AGNs by galaxies; improve the
measurement of $H_ot_o$ to the point that it can distinguish
between low density models with and without $\Lambda$; improve
the constraints on  the amount and distribution of mass from
measurements of galaxy distributions, peculiar motions, and
gravitational lensing; check the theory of structure formation
through X-ray, optical, infrared, and radio surveys of the
evolution of the intergalactic medium, galaxies, and clusters of
galaxies; and maybe even test the dark and $\Lambda$-matter 
hypotheses through advances in particle physics. This work may
yet lead us to an impasse, hypotheses multiplying faster
than the data. That would drive us to new ideas, which would be
exciting. The alternative is that we 
end up with an extensive and compellingly tight network of
tests of a cosmology close to what we have now, which would be
gratifying. 

The second consideration compares two lines of research. In his 
contribution to these Proceedings Neil Turok considers what the
universe might have been like at redshifts so high the 
Friedmann-Lema\^\i tre model certainly could not have applied.
Turok very correctly emphasizes that the question is open and
absolutely must be addressed. But we have to live with the fact
that an empirical validation of the answer may be a 
long time coming. Most papers in these Proceedings deal with the
more limited goal of understanding the large-scale nature of
spacetime and its material content within our Hubble length, now
and back in time through some ten orders of magnitude of
expansion factor. This certainly is not a modest program either,
but the empirical situation is remarkably good: the standard
theory has passed demanding observational tests, and work in
progress promises substantial improvements. The empirical basis
for research on the early universe is a lot more limited. This is
an example of the ways in which the well tested and established
cosmology is incomplete; another is that we can't say what the
dark matter is. But any active physical science is similarly
incomplete: each has a well-tested center around which is the
exciting confusion of ongoing research. 

We all can make pretty good judgments about which elements of 
our subject are well and reliably established, and which are
working hypotheses, when we put our minds to it. And the
community has a reasonably accurate calibration of where each of
us may tend to err on the side of caution or optimism. Our
colleagues in other fields can't be expected to make these calls,
and we shouldn't be surprised that when they have do so they may 
arrive at unduly pessimistic conclusions. We know how to remedy
this, and should put our minds to it. 

I considered cosmology a real physical science decades ago,
though with a meager well-established center. The big recent
change has been the rate of addition to the established center.
But I don't think we're in danger of running out of meaningful
research on open issues any time soon. 

\section{A Next Generation of Cosmological Tests}

Martin Rees comments on the future of research in cosmology once
the present round of tests is satisfactorily concluded. Here I
add some thoughts on another round of cosmological tests of the
physics of the very early universe. 

The rules of evidence in science have evolved to admit 
quite indirect approaches. The community agrees
that the many laboratory tests of quantum
mechanics fully validate it as a real and magnificently
successful physical science, even 
though no one has ever seen a state vector in nature. 
If the CBR revealed a distinctive signature of the tensor
curvature fluctuations predicted in some implementations of
inflation then I think most of us would accept it as an
indirect but strong piece of evidence that inflation really did
happen, even though none of us was there to see it.    

One version of the deconstructionist picture of science as I
read about it is that clever people make up internally
consistent stories to fit agreed-upon conditions, and that
another group could have made up another story, equally
consistent, with an equally satisfactory fit to some similar or
maybe different set of agreed-upon conditions. Those of us who
believe we have convincing evidence physical science describes
aspects of an objectively real world, even on scales very
different from what we can hold in our hands, reply that our
theories have been validated by agreement with tightly
over-constrained and cross checked empirical 
tests. Inflation as we now understand it can be adjusted to 
fit a broad range of possible empirical results. This situation
is unnervingly close to the deconstructionist picture unless 
we stipulate that inflation is a working hypothesis. 

Michael Turner has asked whether empiricists like me would
promote inflation from working hypothesis to established science 
if advances in basic physics produced a unified fundamental
theory that is internally 
consistent, passes all laboratory tests, and predicts fields and
interactions that unambiguously produce inflation. If this
fundamental theory allowed no free parameters to be adjusted to
fit the astronomy, and within the uncertainties of the 
astrophysics it predicted the full suite of observations, it
would be a brilliant addition to established cosmology. A less
perfect fundamental theory might have free parameters, some of 
which could be fixed by laboratory measurements, while others
would have to be determined by the constraints from a cosmology 
established through the rules of evidence one sees applied in
these Proceedings. Should we be satisfied if this theory could be
adjusted to fit all the observations? We would be well
advised to adopt it in most of our analyses of astronomy, but 
not to accept an adjustment of the rules of evidence to admit 
it as the established picture. I suppose most of us think of the
early universe as something that really happened, and things that
happen tend to leave traces. Let us cling to the hope
that something will turn up.

\section{Concluding Remarks}

The evidence assembled in these Proceedings favors the existence
of several kinds of matter: one that acts like
Einstein's cosmological constant, with density parameter 
$\Omega _\Lambda\sim 0.75$, nonbaryonic low pressure matter
with density parameter $\Omega _{\rm DM}\sim 0.2$, baryons with 
density parameter $\Omega _{\rm baryons}\sim 0.05$, and neutrinos
with $\Omega _\nu\sim 0.001$. I believe it is too soon to add
the first number to the list of firmly established elements
of the science of cosmology, because it depends on the model for 
structure formation, and some of us see apparent problems with
the model. People have been discussing the cosmological
parameters for seven decades; we can wait a few more years to
determine  whether we have got the science right.   

If advances in the applications of the cosmological tests firmly
established the values of the fundamental parameters of the
Friedmann-Lema\^\i tre model it would mean general relativity
theory has satisfied demanding tests on the scales of cosmology,
and that we have a well-tested history of 
structure formation. But I would not be surprised to find that
this advance leaves us with the challenge of establishing the
physics of the very early universe.   

\acknowledgements

I thank Mike Disney, Jerry Ostriker, Michael Turner, and Neil
Turok for stimulating discussions. This work was supported in
part by the US National Science Foundation.

\end{document}